\documentclass[
a4paper,
reprint,
twocolumn,
aps,
pre,
amsmath,
assume,
superscriptaddress,
hyperref,
longbibliography
]{revtex4-2} 
\usepackage{graphicx}
\usepackage{dcolumn}
\usepackage{bm}
\usepackage{color}
\usepackage{xcolor}
\usepackage{xfrac}
\usepackage{mathrsfs}
\usepackage{comment}
\usepackage{mathtools}
\usepackage{titlesec}
\usepackage{appendix}
\usepackage{amsfonts}

\newcommand{\new}[1] {\textcolor{black}{ #1} }
\newcommand{\alert}[1] {\textcolor{black}{ #1} }

\def\bea{\begin{eqnarray}}
\def\eea{\end{eqnarray}}
\def\beq{\begin{equation}}
\def\eeq{\end{equation}}
\def\f{\frac}

\def\D{\Delta}
\def\h{\theta}

\def\a{\alpha}
\def\e{\D_{\theta}}

\def\rv{{\bf r}}
\def\v0{{\bf 0}}

\def\la{\langle}
\def\ra{\rangle}
\def\nn{\nonumber}

\def\s{\sigma}

\def\d{\delta}
\def\p{\partial}
\def\l{Pe}

\def\g{\gamma}

\def\a{\alpha}
\def\d{\delta}
\def\p{\partial} 

\def\la{\langle}
\def\ra{\rangle}
\def\e{\epsilon}

\def\g{\gamma}
\def\hf{\frac{1}{2}}

\def\uv{ {\bf u}}
\def\rv{ {\bf r}}
\def\r0{ {\bf 0}}

\def\e{D_\h}
\def\r{\rho}

\begin{document}
\title[]
{Activity-induced phase transition and coarsening dynamics in dry apolar active nematics}

\author{Arpan Sinha}
\email{arpan.s@iopb.res.in}
\affiliation{Institute of Physics, Sachivalaya Marg, Bhubaneswar 751005, India}
\affiliation{Homi Bhabha National Institute, Anushaktinagar, Mumbai 400094, India}

\author{Debasish Chaudhuri}
\email{debc@iopb.res.in}
\affiliation{Institute of Physics, Sachivalaya Marg, Bhubaneswar 751005, India}
\affiliation{Homi Bhabha National Institute, Anushaktinagar, Mumbai 400094, India}

\begin{abstract}
Using the Lebwohl-Lasher interaction for reciprocal local alignment, we present a comprehensive phase diagram for a dry, apolar, active nematic system using its stochastic \new{off-lattice} dynamics. \new{The nematic-isotropic transition in this system is first-order and occurs alongside a fluctuation-dominated phase separation.} Our phase diagram identifies three distinct regions based on activity and orientational noise relative to alignment strength: a homogeneous isotropic phase, a nematic phase with giant density fluctuations, and a coexistence region. Using mean-field analysis and hydrodynamic theory, we demonstrate that reciprocal interactions lead to a density fluctuation-induced first-order transition and derive a phase boundary consistent with numerical results. Quenching from the isotropic to nematic phase reveals coarsening dynamics where nematic ordering precedes particle clustering.  Both the nematic and density fields exhibit similar scaling behaviors, exhibiting dynamic exponents $z_S \approx 2.5$ and $z_\r \approx 2.34$, consistently falling within the range of 2 and 3.
%
\end{abstract}

\maketitle

\section{\label{sec:level1}Introduction}

Active nematics represent a fascinating class of nonequilibrium systems characterized by the collective motion and organization of self-propelled entities~\cite{Simha2002, Ramaswamy2003, Ramaswamy2010, Marchetti2013, Ramaswamy2019, Shaebani2020, Bar2020}. Inspired by a myriad of biological systems, ranging from the dynamic organization of the cytoskeleton within cells to the emergent behaviors observed in tissues and even granular materials, active nematics have emerged as a pivotal area of study at the interface of physics, biology, and materials science~\cite{Julicher2007, Prost2015, Bechinger2016, Duman2018, Joshi2019, Redford2024, Ansari2022, DeCamp2015, Shi2010, Shi2014, Bertin2013a}.
At the heart of active systems lies the concept of self-propulsion, where individual constituents continually convert internal energy into directed motion. This intrinsic activity, prevalent in biological systems such as motile cells and bacterial colonies, fuels a rich array of dynamic phenomena, including the spontaneous formation of orientational order, collective motion, and the emergence of complex spatiotemporal patterns~\cite{Marchetti2013, Julicher2007, Prost2015}.
Moreover, active nematics find analogs in nonliving systems, including vibrated granular matter and artificial microswimmers suspended in fluids. The last few decades have seen tremendous progress in understanding their collective properties and phase behaviors~\cite{Simha2002, Ramaswamy2003, Ramaswamy2010, Marchetti2013, Bar2020, Shi2010, Shi2013, Shi2014, ngo2014large, chate2006simple, Bhattacherjee2019, Moore2021}.

In dry active matter, self-propelled particles (SPP) with local alignment display order-disorder transition coupled with density fluctuations. Consideration of ferromagnetic alignment of the active heading directions of individual propulsion led to flocking and persistent motion of the flocks. This was demonstrated in the celebrated Vicsek model and its variants and described by the Toner-Tu theory of coupled orientation and density fields~\cite{vicsek1995novel, Gregoire2004, Chate2008, Toner1995a, Marchetti2013}. 
\alert{Self-propelled rods with nematic alignemnt~\cite{Kraikivski_2006, Peruani2006, Baskaran_2008, Peshkov_2012, grossmann2016mesoscale, Bar2020} and recent experimental studies~\cite{Huber_2018} display an isotropic-to-nematic transition,  with the formation of nematic bands first observed in Vicsek-like models of polar active nematics~\cite{ginelli2010large}. Numerical studies of self-propelled rods~\cite{Peruani2006, ginelli2010large, Grossmann_2020} as well as experiments on {\em M. xanthus} in Ref.~\citenum{starruss2012pattern} showed the formation of polar clusters. The overall polar order vanishes in large-scale simulations of related models that led to long-range nematic order in 2D~\cite{ginelli2010large}. Recent experiments in actomyosin motility assays~\cite{Huber_2018} indicate that both polar or nematic order can emerge depending on control parameters.}
%
%
%
A Boltzmann-Landau-Ginzburg kinetic theory approach, based on a Vicsek-like model of active nematics, produced hydrodynamic equations consistent with previous findings~\cite{Ramaswamy2003, Ramaswamy2010, Marchetti2013, Peshkov_2012, Shi2010, Shi2014, Bertin2013a}.
In certain systems, alignment results from physical processes such as actual collisions between active elements~\cite{Blair2003, Peruani2006,
 Narayan2007,Duman2018, Joshi2019, Ansari2022}. Other instances involve effective interactions mediated by non-equilibrium mechanisms like complex biochemical signaling or visual and cognitive processes~\cite{chepizhko2021revisiting, Helbing2000, Saha2020}. The former alignments are characterized by reciprocal interactions, while non-reciprocal rules often describe the latter~\cite{Ivlev2015, Fruchart2021, Loos2022}.
In equilibrium systems, microscopic details are typically considered irrelevant to emergent macroscopic properties if the models share symmetries and conservation laws. However, recent studies on aligning active particles with short-range interactions, be it vectorial or nematic active matter, revealed that microscopic implementations in the form of reciprocity and additivity can influence macroscopic behaviors~\cite{chepizhko2021revisiting, Sinha2024}. For example, active apolar nematics in the presence (or absence) of reciprocity show first-order (or continuous) nematic-isotropic (NI) transition in the presence of fluctuation-dominated phase separation (FDPS) of high and low-density regions~\cite{Sinha2024}. \new{These findings align with the non-reciprocal Vicsek-like models~\cite{chate2006simple, ngo2014large}, exhibiting an unimodal distribution of nematic order across the transition, indicative of a continuous NI transition, despite in the presence of a coexistence of low-density regions with unstable high-density nematic bands.}

SPPs can lose their overall polarity when there are rapid reversals in the self-propulsion direction. In active nematics, particles spontaneously align along an axis $\hat{\bf n}$ with a $\hat{\bf n} \to -\hat{\bf n}$ symmetry~\cite{Simha2002, Ramaswamy2003, Bertin2013a, Mishra2006, chate2006simple, ngo2014large, Das2017a}. Examples of such systems include colliding elongated objects~\cite{Peruani2006, ginelli2010large, Peruani2012}, migrating cells~\cite{Gruler1995, gruler1999nematic}, cytoskeletal filaments~\cite{Balasubramaniam2022}, certain direction-reversing bacteria~\cite{Wu2009, Theves2013a,starruss2012pattern, Barbara2003, Taylor1974a}, and vibrated granular rods~\cite{Blair2003, Narayan2007}. 
Even at high activity levels, the apolar nature of the particles results in zero macroscopic velocity. Intriguingly, the collective properties of these apolar SPPs still show a clear dependence on activity, as orientational fluctuations drive particle current~\cite{Ramaswamy2003, Mishra2006, Shi2010, Shi2014, Bertin2013a, ngo2014large}.

In this paper, we consider a dry, apolar, active nematics model using a reciprocal additive alignment interaction between local neighbors. 
\alert{This model is idealized; real systems with active polymers, F-actins, or microtubules may display additional properties, such as emergent elasticity or chiral segregation, which are not covered here~\cite{Redford2024,Duman2018, Joshi2019, Moore2021}.} 
An earlier on-lattice implementation of active Lebwohl-Lasher model~\cite{Lebwohl1972} found an FDPS associated with the ordering transition~\cite{Mishra2006}. Our earlier off-lattice calculation of SPPs at a fixed activity demonstrated a clear first-order NI transition with increasing orientational noise~\cite{Sinha2024}. 

 In this paper, we present a comprehensive phase diagram as a function of changing activity and relative orientational noise with respect to the alignment strength. 
\alert{Previous studies on SPP often used density and noise as control parameters~\cite{Shi2014, ngo2014large}. This work emphasizes the recurring theme of order emerging from activity~\cite{Kraikivski_2006, Baskaran_2008}, demonstrating how the strength of activity influences the phase diagram.}
 It shows three different parameter regimes characterized by (i)~a homogeneous isotropic phase, (ii)~a nematically aligned phase with significant density fluctuations, and (iii)~a coexistence of nematic and isotropic in the presence of the largest density fluctuations. The phase transition is characterized by FDPS and first-order NI transition across all parameter regimes. Using a mean-field hydrodynamic argument, we obtain an analytic prediction for the phase boundary that shows reasonable agreement with our numerical results. Our primary contribution is this direct calculation of the detailed phase diagram. 

Moreover, we study the phase ordering dynamics after a deep quench from isotropic to nematic phase. The stochastic simulations show dynamical scaling for density and nematic fields, with similar scaling exponents with values bounded between non-conserved and conserved dynamics consistent with earlier hydrodynamic calculations~\cite{mishra2014aspects}. We observe a delay in particle clustering dynamics relative to nematic ordering, with the delay time controlled by the depth of the quench.  Our second main contribution is the direct calculation of phase ordering using stochastic simulations.

The rest of the paper is structured as follows: Section~\ref{Model} details the model of dry active nematic with reciprocal alignment. Section~\ref{phase_diagram} presents the comprehensive phase diagram. Section~\ref{sec_mft} describes the mean field hydrodynamic analysis, explaining the phase boundaries and the nature of the NI transition. We then explore the coarsening of nematic order and density field in Section~\ref{phase_ordering}. Finally, in Section~\ref{conclusion}, we conclude by summarizing our main results and presenting an outlook.

\section{ Model}\label{Model}
{
In this study, we explore how $N$ dry active apolar particles align collectively in a nematic manner in a two-dimensional (2D) area $A=L\times L$. With a fixed active speed $v_0$, we describe the particles' microstate as $\{ \rv_i, \uv_i, q_i \}$, where $\rv_i$ denotes position, $\uv_i$ signifies velocity, and $q_i$ represents orientation. The evolution of particle positions unfolds as:
\bea 
\rv_i(t+dt) = \rv_i(t)+ q_i\, v_0 \uv_i\, dt.
\label{eq_dr}
\eea
We adopt periodic boundary conditions. The polarity $q_i=\pm 1$ is randomly assigned with equal probability, \alert{at each time step} to model apolar particles. The heading direction $\uv_i = (\cos \h_i, \sin \h_i)$ evolves along with the angle $\h_i$ with respect to the $x$-axis. Dynamics are governed by a competition between inter-particle alignment and orientational noise. In active nematics, alignment interactions are designed to make neighboring particles' heading directions parallel or anti-parallel with equal likelihood.

We examine the interaction of particle heading directions using the Lebwohl-Lasher potential~\cite{Lebwohl1972, Mishra2006}:
\bea
U = -J \sum_{\la ij\ra} \cos [2 (\h_i -\h_j)]
\label{eq_LL}
\eea
within a cutoff distance $r_c$ determining the interaction range $r_{ij}=|\rv_i - \rv_j| < r_c$, and setting the unit of length.  Following the Ito convention, the orientational Brownian motion of $\uv_i(\h_i)$ is governed by:
\bea
\h_i(t+dt) = \h_i(t) -\mu (\p U/\p \h_i) dt + \sqrt{2 D_r}\, dB_i(t).
\label{eq_model_1}
\eea
Here, $\mu$ represents mobility, $D_r$ is the rotational diffusion constant, and $dB_i$ is a Gaussian process with zero mean and correlation $\langle dB_i dB_j\rangle=\delta_{ij} dt$. These equations depict persistent motion for a free particle, where $D_r$ determines the persistence time $\tau_p = D_r^{-1}$, setting the unit of time. The model illustrates apolar particles aligning nematically with strength $J>0$. It is important to note that the torque experienced by a particle pair $i,j$ is equal and opposite to each other.

Employing the Euler-Maruyama scheme, we perform direct numerical simulations by integrating Eq.(\ref{eq_dr}) and (\ref{eq_model_1}) using $dt=7 \times 10^{-4}\, D_r^{-1}$. Care is taken so that $v_0 dt < r_c$ to incorporate the influence of interaction properly\alert{; the condition ensures that interaction between particles is well resolved in the dynamics~\cite{footnote}.
} 
We utilize the scaled angular diffusion $\e:= {2 D_r}/{\mu J}$ and activity \alert{in terms of P{\'e}clet number} $Pe:= v_0/D_r r_c$ as \alert{dimension-less} control parameters to construct the phase diagram, delineating associated phases at a fixed dimensionless number density $\rho={N}/{L^2}=\r_0$ with $\r_0 r_c^2=0.4$. \alert{The scaled angular diffusion $\e$ is a ratio of orientational relaxation time $(\mu J)^{-1}$ due to alignment interaction to the single-particle persistence time $D_r^{-1}$. The P{\'e}clet number $Pe$ is defined as a ratio of active persistence length $v_0/D_r$ of a single SPP to the alignment interaction range $r_c$, to capture the effect of interaction. }

 The nematic order parameter is denoted as 
 \(\langle Q_{\alpha\beta} \rangle\), where 
\(
Q_{\alpha\beta}^{(i)} =\left(u_{i\alpha} u_{i\beta} - \frac{1}{d} \delta_{\alpha\beta}\right)
\)
for the \(i\)-th particle, with \(\alpha\), \(\beta\) denoting the component indices. 
\alert{The averaging $\la \dots \ra$ in \(\langle Q_{\alpha\beta} \rangle\) is computed over all particles within the region of interest, which could be the entire system for the system-wide order parameter or a local coarse-graining volume, and averaged across steady-state configurations.} In 2D,
\[
{\bf Q}^{(i)} =\f{1}{2}  
\begin{pmatrix}
\cos 2\theta_i & \sin 2\theta_i \\
\sin 2\theta_i & -\cos 2\theta_i
\end{pmatrix}.
\]
\alert{From numerical simulations, we determine the degree of nematic order by analyzing the positive eigenvalue of $\la Q_{\alpha\beta} \ra$, which represents the scalar order parameter~\alert{(see Appendix-\ref{2dorder} for further details)}}:
\bea
S = \left[\langle \cos(2\theta_i) \rangle^2 + \langle \sin(2\theta_i) \rangle^2\right]^{1/2}.
\eea
  The value of $S$ is bounded on both sides, $0 \leq S \leq 1$. 
To compute the steady-state average $S$ across the entire system, we average over all $N$ particles and further average over steady-state configurations, represented by red line points in Fig.\ref{fig:phase_diag}(b). Additionally, we determine the standard deviation of the global order parameter $S$ and identify its maximum as the transition point $\e^{\ast}$ for a given $Pe$; these are indicated by blue line points in Fig.\ref{fig:phase_diag}(b). For a localized assessment of $S$, we perform averaging over particles within a local volume. This is used to determine $S$ over coarse-grained volumes of linear dimension $10 r_c$ to calculate the probability distributions $P(S)$ in Fig.\ref{fig:phase_diag}(c).

\begin{figure*}[t]
\centering
  \includegraphics[width=\linewidth]{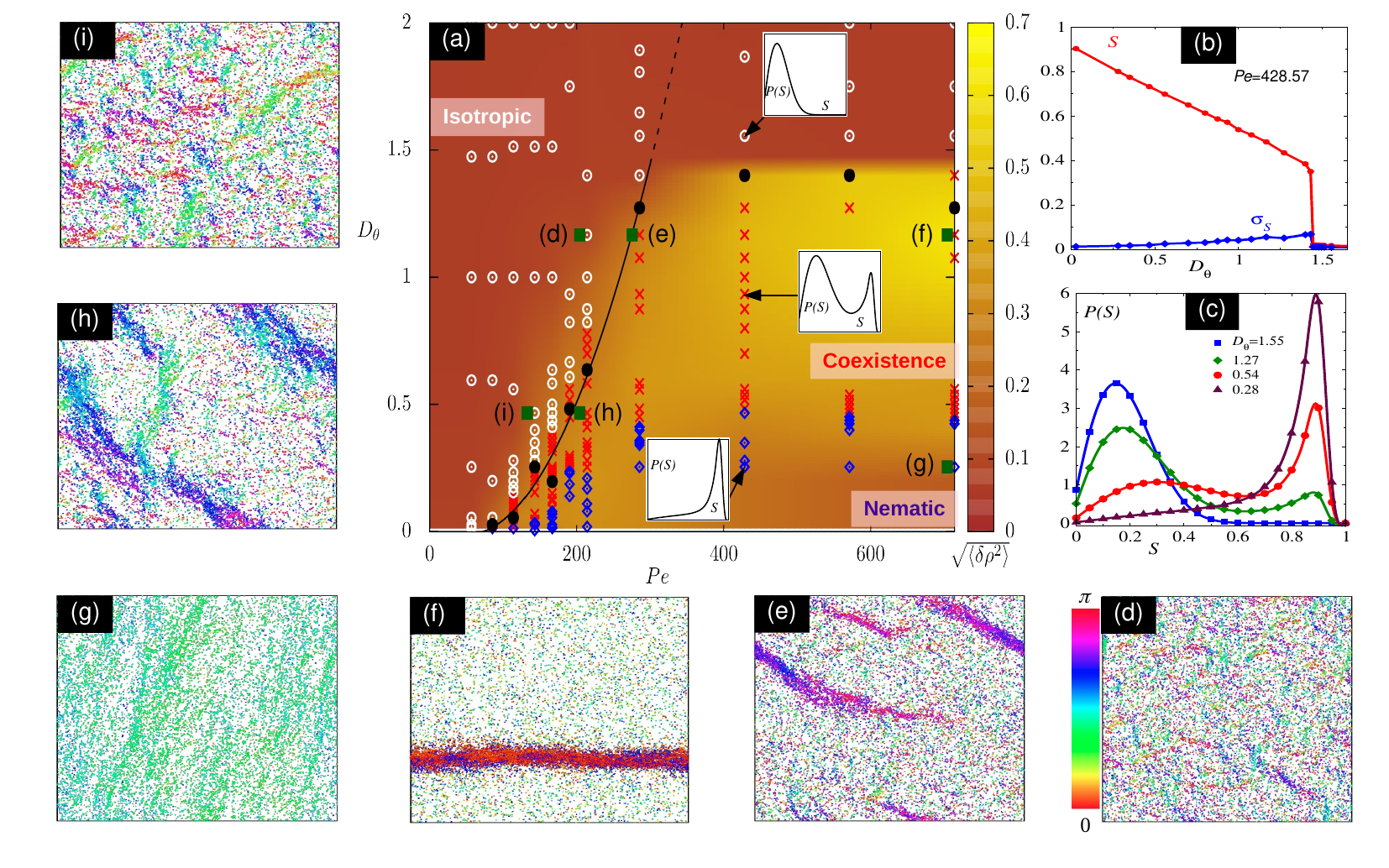}
  \caption{(a)~Phase diagram in $(Pe,\e)$ plane with dimensionless activity $Pe=v_0/D_r r_c$ and orientational noise relative to alignment strength $\e= 2 D_r/\mu J$ showing three distinct regions: \alert{isotropic homogeneous~(Isotropic), nematic-isotropic coexistence~(Coexistence) and nematically ordered~(Nematic).} 
  The solid black line has the scaling form $\e^{*}\sim Pe^2$~(see Sec.\ref{sec_mft} for derivation). (b)~The nematic order $S$ (red line points) and standard deviation $\s_s$ (blue line points) as a function of $D_{\theta}$ at a fixed $\l=428.57$. (c)~The local probability distribution $P(S)$ at $\l=428.57$ and various $\e$ mentioned in the figure legend. The distributions are calculated using $S$ calculated over local coarse-grained volumes of linear dimension $10\,r_c$.  \alert{(d-i)}~Representative configurations at parameter values marked by solid green boxes in the phase diagram in (a). The parameter values are $~(Pe,\e)=$ \alert{(214.28, 1.166), (285.71, 1.166), (714.28, 1.166), (714.28, 0.254), (214.28, 0.466), (142.85, 0.466) respectively.} The color palette to the left of Figure (d) denotes the heading directions of the particles in all the snapshots shown here.  } 
  \label{fig:phase_diag}
\end{figure*}    
 
\section{Phase diagram}\label{phase_diagram}

We begin our investigation by studying the steady states of the model in square geometry of size $L=200$ and $N=16000$, utilizing periodic boundary conditions (PBC). 
Fig.\ref{fig:phase_diag}(a) shows the phase diagram of the NI transition.
The regions of the nematic phase (\alert{Nematic})  are denoted by blue diamonds, the region of NI coexistence (\alert{Coexistence}) is indicated by red crosses, and the homogeneous isotropic phase (\alert{Isotropic}) region is marked by white open circles. The phase boundary between Isotropic and Coexistence is denoted by black-filled circles. 
\alert{The phase diagram clearly demonstrates that nematic order arises from activity: the system remains in an isotropic phase at $Pe=0$, and only when $Pe$ exceeds a certain threshold does nematic order begin to emerge.
}
The heat map in the phase diagram shows the amount of density fluctuations. This is calculated in coarse-grained volumes of linear dimension $10r_c$ using $\d \r = [\la \r^2\ra - \la \r\ra^2]^{1/2}$, and then averaging over the system size and all steady state configurations. As the color codes show, the density fluctuation is maximum in the coexistence. Its presence is significant even in the ordered nematic phase. The fluctuation is minimal in the homogeneous isotropic phase.

The three insets in Fig.\ref{fig:phase_diag}(c) illustrate the typical distributions of the scalar nematic order parameter, corresponding to nematic, coexistence, and isotropic regions.
An analytic description of the NI phase boundary $\e \sim \l^2$ is provided in the following section. Shown by a black solid line in Fig.\ref{fig:phase_diag}(a), it qualitatively captures the actual phase boundary up to $\l \approx 300$.  The broken solid line indicates a breakdown of this simple estimate. Beyond this point, the phase boundary becomes approximately independent of $\l$.  
\alert{Unlike in equilibrium systems, where number fluctuations follow \(\langle \Delta n^2 \rangle = \langle n^2 \rangle - \langle n \rangle^2 \sim \langle n \rangle^a\) with \(a = 1\), active nematic systems exhibit \(a > 1\), known as giant number fluctuations~\cite{Simha2002,Sinha2024,Mishra2006}. Consistent with this expectation, we observe \(a \approx 1.52\) in the nematic phase at \(Pe = 714.28\) and \(D_{\theta} = 0.254\); details can be found in Appendix-\ref{app_gnf}.
Additionally, as previously demonstrated~\cite{Sinha2024}, the active nematic model employing the Lebwohl-Lasher potential exhibits FDPS~\cite{Das2000}.
}
%

}

The NI transition is captured using the dependence of the scalar order parameter $S$ and its standard deviation $\s_s = [\la S^2\ra - \la S\ra^2]^{1/2}$ on $\e$, at a fixed $\l$; see Fig.\ref{fig:phase_diag}(b). 
The cross-correlation between fluctuations in nematic order and density peaks near the transition, as detailed in Appendix-\ref{app_xcor}. However, this correlation diminishes significantly within the nematic phase and approximately vanishes after transitioning to isotropic.
The local probability distribution $P(S)$ in Fig.\ref{fig:phase_diag}(c) shows an unimodal distribution with the maximum at small $S$ for large $\e$, transforms into a bimodal distribution at intermediate $\e$ values. This clearly indicates a phase coexistence, characterizing a first-order transition. At an even smaller $\e$, the distribution becomes unimodal again, this time displaying a single maximum at large $S$, indicating a nematic phase.   

Representative configurations are shown in Fig.\ref{fig:phase_diag}(d)-(i), with orientations distinguished by a color palette shown on the left of Fig.\ref{fig:phase_diag}(d). The parameter values associated with these configurations are marked by solid green squares on the phase diagram Fig.\ref{fig:phase_diag}(a).
A typical configuration of the isotropic phase in Fig.\ref{fig:phase_diag}(d) shows an approximately homogeneous distribution of particles and their random orientations of heading direction. At a higher $\l$, in Fig. \ref{fig:phase_diag}(e), we observe a local clustering of particles in oriented bands. These bands display a higher nematic order. At an even higher $\l$ in Fig.\ref{fig:phase_diag}(f), simulations show system-spanning bands with large nematic order coexisting with a uniform isotropic background. This corresponds to the bimodal distribution in $P(S)$ shown in Fig.\ref{fig:phase_diag}(c). However, the nematic bands show transverse instability; they form and disappear~\cite{putzig2014phase, Mishra2006, Shi2013, Shi2014, grossmann2016mesoscale}. Appendix \ref{nematic_bands} presents configurations as bands form and break. Decreasing $\e$, keeping $\l$ unchanged, in Fig.\ref{fig:phase_diag}(g), we observe nice nematic order all through the system, corresponding to a homogeneous nematic fluid phase. The corresponding distribution $P(S)$ is unimodal with the maximum at large $S$. Again, in Fig.\ref{fig:phase_diag}(h), we show a typical configuration at a smaller $\l$ and $\e$, near the phase boundary, to find local nematic bands coexisting with an isotropic background. At an even smaller $\l$, the system gets back to the isotropic phase in the presence of large density fluctuations; see Fig.\ref{fig:phase_diag}(i).

\section{Mean field and hydrodynamic analysis}\label{sec_mft}
Here, we consider the coupled evolution of the nematic order parameter density $\Pi_{ij}(\rv,t)=Q_{ij}(\rv,t)\rho(\rv,t)$ and particle density $\r(\rv,t)$~\cite{Shi2010, Shi2014, Bertin2013a, Marchetti2013}.  
The local field of nematic order
\begin{align}
   {\bf Q}(\rv,t)=& \f{S(\rv,t)}{2} 
   \begin{pmatrix}
  \cos 2 \theta (\rv,t) & \sin 2 \theta (\rv,t) \\
  \sin 2 \theta (\rv,t) & -\cos 2 \theta (\rv,t) 
  \end{pmatrix} 
  \label{eq_Qcg}
\end{align}
is determined by the local scalar order $S(\rv,t)$ and orientation $\h(\rv,t)$.

\alert{In the following, we present a mean-field description of the coupled evolution of the density $\r(\rv, t)$ and order parameter field $Q(\rv,t)$ for active apolar nematics~\cite{Marchetti2013, Ramaswamy2003, Shi2010, Bertin2013a}.}

\alert{Local curvature in the nematic field can induce a temporary polarity, which, when activity is present, drives a current in the direction determined by this polarity~\cite{Marchetti2013}. This results in an active current given by $J^{(a)}_i = -\l\, \p_j \Pi_{ji}$.}
\alert{Using the active current $J^{(a)}_i$ and the diffusive current $ J^{(d)}_i = -D \p_i \rho$, the evolution of particle density field, governed by the continuity equation $\p_t \r = -\p_i (J^{(a)}_i+J^{(d)}_i)$ leads to,} 
\bea
\p_t \rho =  \l \,\p_i \p_j \Pi_{ji} + D \nabla^2 \rho, 
\label{eq_rho}\eea
where $D$ denotes the diffusion constant. 
The evolution of nematic order follows~\cite{Marchetti2013, Ramaswamy2003, Bertin2013a}, 
\bea
 \p_t \Pi_{ij} = [\a_1 - \a_2\, {\rm Tr}(\Pi^2) ] \Pi_{ij}  + D \nabla^2 \Pi_{ij} 
 + \dfrac{\l}{4} \left[\p_i \p_j - \hf \d_{ij} \nabla^2\right]\rho \, .
 \label{eq_pi}
\eea
\alert{In the above equation, the term $[\a_1 - \a_2\, {\rm Tr}(\Pi^2) ] \Pi_{ij}$ 
can be derived from a Landau-de Gennes free energy functional, which effectively describes an equilibrium continuous transition between the nematic and isotropic phases in 2D~\cite{degennes_book, pcmp1995}.
For a traceless symmetric tensor \(\Pi\), the conditions \( \text{Tr}(\Pi^3) = \text{Tr}(\Pi) = 0 \) in 2D imply that cubic order terms in \(\Pi\) are not present in the free energy functional. Consequently, this leads to the absence of quadratic order terms in \(\Pi\) in the equation above.
As we show in Appendix-\ref{app_eqmft}, the coefficients in the Landau-de Gennes theory are given by $\a_1 = [1-{D_\h}/{\e^{(c)}} ] $ and $\a_2 = ({1}/{4})[{\e^{(c)} }/{D_\h}]^2$ with the mean-field critical point $\e^{(c)}  = A \r$. 
The $D \nabla^2 \Pi_{ij}$ term in the above equation is the Ginzburg-like term in the traditional Landau-Ginzburg description~\cite{pcmp1995}. 
The last term in Eq.\eqref{eq_pi}, which couples the evolution of nematic order to density field gradients, arises solely from the active processes~\cite{Marchetti2013, Bertin2013a}.}

\alert{The interplay between the conserved density field and the non-conserved order-parameter field resembles Model C from the Hohenberg-Halperin classification~\cite{hohenberg1977theory}. However, when activity is absent (Pe = 0), the evolution of density and nematic order within our model gets decoupled. Consequently, our phase diagram and the coarsening dynamics in density have no equilibrium counterpart.}

\begin{figure*}[t]
\centering
  \includegraphics[width=\linewidth]{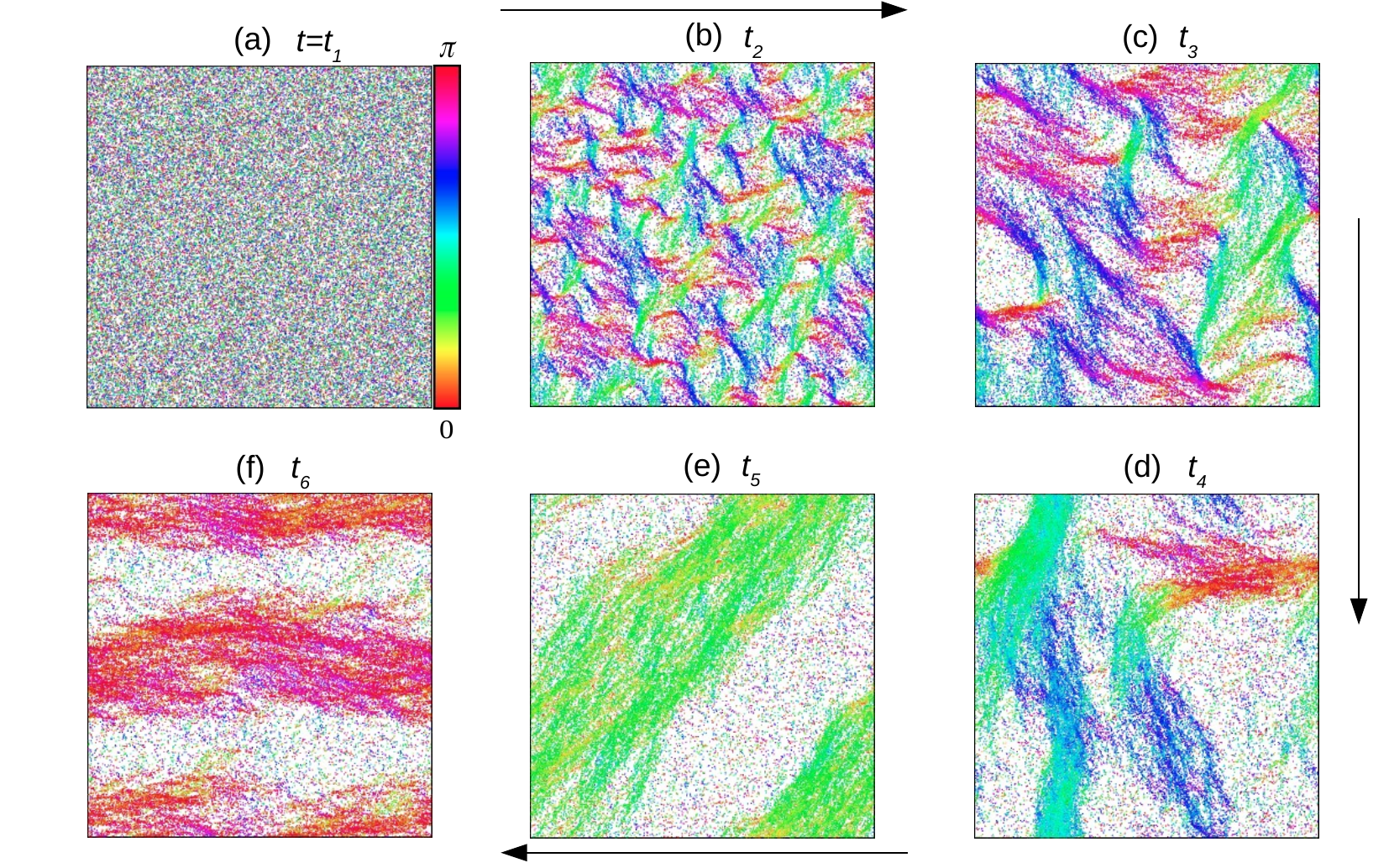}
   \caption{Typical configurations corresponding to the evolution after a quench from $\e=2.8$ to $\e=0.28$ at $\l=428.57$ are shown. The color palette in (a) encodes orientations of particle heading directions. The black arrows show the direction of increasing time. At initial times, $t_1=0.007 \, D_r^{-1}$, the system is in a homogeneous disordered phase~(a). By $t_2=2.8\, D_r^{-1}$, thin nematic bands of higher density are formed. They are oriented in multiple directions~(b). At a later time, $t_3=21\, D_r^{-1}$, these bands coarsen into thicker and longer branched patterns~(c). At an even later time $t_4=350\, D_r^{-1}$, they coarsen into broader and system-spanning bands~(d). By the time $t_5=12600 \, D_r^{-1}$, steady-state system-spanning nematic bands appear~(e). These bands break, form, and orient along spontaneously chosen directions that vary with time. For example, see another steady-state configuration at $t_6=19600 \, D_r^{-1}$~(f). Here we use $N=64000$ particles,  $Pe=428.57$, $\rho_0 r_c^2=0.4$, and $L=400 r_c$.}
\label{fig:coarsening_confs}
\end{figure*}

 {
\alert{Within the mean-field approximation of uniform density and order,}  Eq.\eqref{eq_pi} gives $\f{d Q_{ij}}{dt} = [\a_1 - \a_2 {\rm Tr} (Q^2)]Q_{ij}$ leading to a continuous transition at $\a_1=0$, \alert{i.e., at $\e=\e^{(c)}$}.
Describing the first-order transition \alert{in the nematic order} requires incorporating the \alert{activity induced} density fluctuations~\cite{Chen1978, Halperin1974}, for which we use a renormalized mean field theory~\cite{Tu2019, Das2017a}.
We consider a small activity-induced density fluctuation $\d \rho$ such that  $\rho=\rho_0+\d \rho$ to express $\a_{1,2}(\rho) = \a_{1,2}^0+{\a'}_{1,2}^0 \d \rho$ where $\a_{1,2}^0 = \a_{1,2}(\r_0)$ and ${\a'}_{1,2}^0 = (\p \a_{1,2}/\p \r)_{\r_0}$. The requirement of a zero current steady state in density evolution suggests $\d \r \approx \l \,S/D$~\cite{Sinha2024}. Using this in the mean-field limit of Eq.\eqref{eq_pi}, we obtain the evolution for scalar order
\bea
\p_t S = [u_2 + u_3 S - u_4 S^2] S,
\eea
where, \alert{for notational simplicity we used} $u_2=\a_1^0$, $u_3={\a'}_1^0 \l/D$ and $u_4= \a_2^0$.
This equation can be expressed as a free energy minimizing kinematics $\p_t S = - \p {\cal F}/\p S$ with 
\bea
{\cal F}=-\f{u_2}{2} S^2 - \f{u_3}{3} S^3 +\f{u_4}{4} S^4, 
\eea
keeping up to $S^4$ order term. 

\begin{figure*}[t]
\centering
 \includegraphics[width=0.9 \linewidth]
 {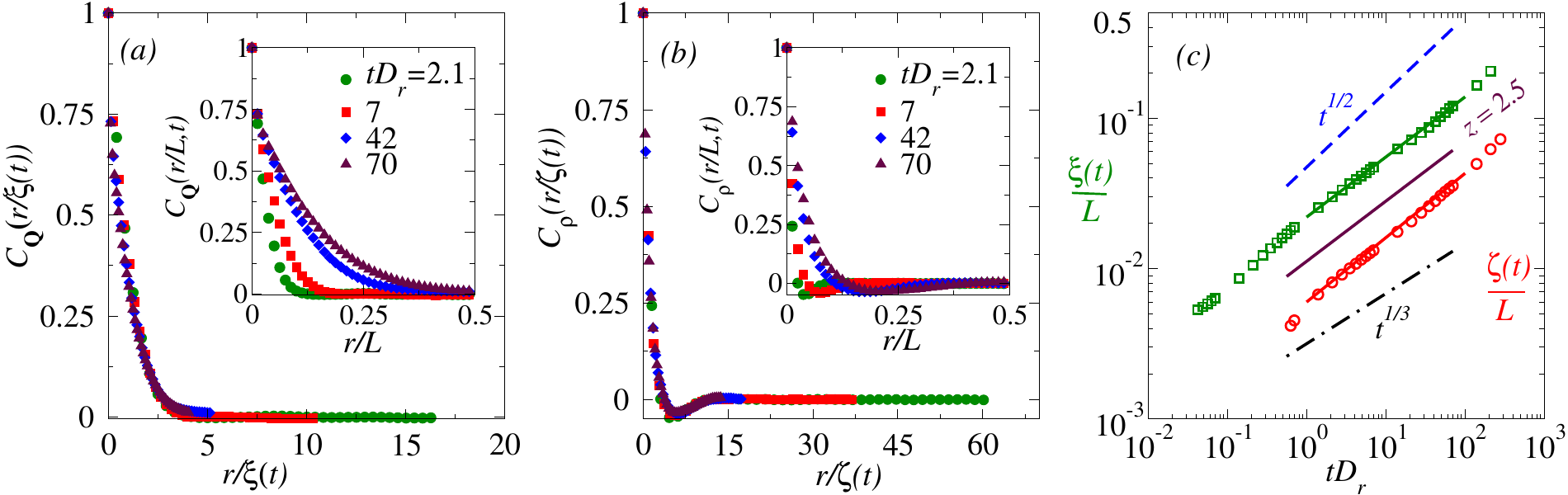}
  \caption{The nematic and density autocorrelation $C_{\bf Q}(r)$~($a$) and $C_\r(r)$~($b$) during quench from $\e=2.8$ to $\e=0.28$ at $\l=428.57$. Insets show the correlations plotted against $r/L$. The main plots show data collapse by scaling separation $r$ by dynamic correlation lengths $\xi(t)$~($a$) and $\zeta(t)$~($b$).  
 $(c)$~The correlation lengths $\xi(t)$~($\square$) for nematic order, $\zeta(t)$~($\circ$) for density exhibit similar power-law growths $\sim t^{1/z}$ in the scaling regime with exponents \alert{$z_S \approx 2.5 \pm 0.03$ and $z_\r = 2.34 \pm 0.02$~(fits are shown using lines through data), values not too different from the numerical estimate of $z_S \approx 2.5$ for equilibrium Lebowhl-Lasher model as indicated by the solid maroon line. The model A growth law $t^{1/2}$~(dashed blue line) and the model B growth law $t^{1/3}$~(dash-dotted black line) are shown for comparison.} 
  Cells of linear dimension $5\,r_c$ are used to calculate the coarse-grained correlations from numerical simulations.  
  }
  \label{fig:Lc_rho_S}
\end{figure*}

The cubic term in $S$ in the effective free energy density ${\cal F}$ results in a first-order NI transition. At the transition, the scalar order parameter jumps from $S=0$ to 
\bea
S_c = \f{2}{3} \f{{\a'}_1^0}{\a_2^0} \f{\l}{D} .
\eea
The transition point changes with $\l$ to give the transition line 
\bea
D_\h^{\ast}
= D_\h^{(c)} \left[1
+ \f{2}{9}\f{({\a'}_1^0)^2}{ \a_2^0} \left(\f{\l}{D} \right)^2\right],
\eea
increasing quadratically with $\l$. A plot of $D_\h^{\ast} \sim \l^2$ using the black solid line in Fig.\ref{fig:phase_diag}(a) approximately captures the phase boundary up to $\e \approx 1.4$. At higher $\e$, even the activity-induced increased density fluctuations fail to sustain the nematic order, rendering the phase boundary independent of $\l$. In fact, as can be observed from the heat map in Fig.\ref{fig:phase_diag}(a), the density fluctuation gets suppressed at these $\e$ values. We note that the first-order NI transition is purely active; it vanishes in the limit of $\l \to 0$ with the vanishing of $S_c$.

Before closing this section, certain comments are in order. Within the Lebowhl-Lasher model used here, the first-order isotropic-nematic transition precludes the possible continuous transition as $\e^\ast > \e^{(c)}$. Moreover, the observed first-order transition relies crucially on the density dependence of $\e^{(c)}$, which resulted from the reciprocal and additive nature of interaction implemented in the Lebowhl-Lasher model. If $\e^\ast$ is independent of density, as in certain non-reciprocal models, the NI transition can become continuous~\cite{Sinha2024}. 

\alert{ 
Incorporating spatial variations through the hydrodynamic equations \ref{eq_rho} and \ref{eq_pi} for apolar nematics helps explain the band formation observed in the coexistence region~\cite{Bertin2013a, grossmann2016mesoscale}.   Subcritical bifurcations of band solutions in direction-reversing self-propelled rods were demonstrated in Ref.~\citenum{grossmann2016mesoscale}.
The term $\p_i \p_j \Pi_{ij}$ is crucial for the development of density inhomogeneities~\cite{Shi2010}.
Similar band formations have been noted in polar nematics and analyzed with hydrodynamic equations~\cite{Peshkov_2012}. }

 \section{Phase ordering kinematics} \label{phase_ordering}
  Having established the phenomenology of the NI transition, we now aim to understand the coarsening kinematics of the nematic order. Towards this end, we present our numerical results of the growth kinetics following a deep quench at a fixed $\l=428.57$ from a homogeneous isotropic initial state at $\e=2.8$ to $\e=0.28$ corresponding to a steady-state nematic. For better statistics, we use a larger system with $N=64000$. Fig.\ref{fig:coarsening_confs} shows a typical series of snapshots depicting this evolution. As time progresses, domains of higher order parameters and higher density form, with their typical size increasing over time. The deep quench leads to local instability. In the beginning, the homogeneous isotropic state shows instability through the formation of nematically ordered filaments crisscrossing each other; see Fig.\ref{fig:coarsening_confs}(b). The small filaments merge and coarsen with time, as shown in Fig.\ref{fig:coarsening_confs}(c) and (d). At a late time, one finds dynamic system-spanning bands of nematically ordered regions; see Fig.\ref{fig:coarsening_confs}(e) and (f). In the steady state, the local undulations in nematic bands make them unstable toward breaking, and one finds repeated formation, breaking, and reorientations of such bands with time~\cite{ngo2014large, Shi2010, Shi2014}.

To quantify these observations, we track the spatial nematic order parameter autocorrelation $C_{\bf Q}(r,t)$  and the density autocorrelation  $C_\r(r,t)$ as the system relaxes towards its final ordered state. To obtain a direct comparison against earlier field theory calculations, we divide the simulation volume into cells of length $\ell_c$. We employ the coarse-grained order parameter ${\bf Q}(\rv,t) = \f{1}{n(t)} \sum_{i=1}^{n(t)} {\bf Q}^{(i)}$ by performing instantaneous averaging over local cells at positions $\rv$ containing $n(t)$ number of particles. 
Similarly, in the same cells, we use a coarse-grained local density $\r(\rv,t)=  n(t)/\ell_c^2$.
We use $\ell_c=5 r_c$ in the numerical calculations.
The correlation functions are defined as follows:
 \bea
C_{\bf Q}(r,t)= \f{Tr\la {\bf Q}(r,t) {\bf Q}(0,t)\ra}{Tr\la {\bf Q}^2(0,t)\ra} ,
 \eea 
and
 \bea
C_\r(r,t)=\f{\la \delta\r(r,t)~\delta \r(0,t)\ra}{\la\delta\r^2(0,t)\ra}
 \eea
 where $\delta\r(r)=\r(r)- \rho_0$ and $r$ denotes the separation between local coarse-grained volumes. For further details, see Appendix-\ref{app_corr1}.

The time-dependent correlation functions plotted in Fig.\ref{fig:Lc_rho_S} are calculated averaging over forty independent trajectories. The simulations show exponentially decaying correlations up to $tD_r\approx 500$. Beyond that, the finite-size effect kicks in; the deep quench renders quasi-long ranged nematic order with power-law decay of correlations and system-spanning bands. At shorter times, the correlation length $\ell$ can be obtained by fitting the correlation functions with $\exp(-r/\ell)$ form.   
 Rescaling the separations by correlation lengths $\ell=\xi(t)$ and $\ell=\zeta(t)$ for nematic and density correlations, we obtain nice data collapse for both the correlation functions, shown in the main figures of Fig.\ref{fig:Lc_rho_S}($a$) and ($b$).  

In the intermediate time scaling regime, both the correlation lengths,  $\zeta(t)$ and $\xi(t)$, increase algebraically.  Note that the building up of orientational correlation precedes that of local density; see the delay in growth of $\zeta(t)$ compared to $\xi(t)$ in Fig.\ref{fig:Lc_rho_S}($c$). This delay is caused by the time required for nematic instability to influence the density fluctuation. As can be seen from Eq.\eqref{eq_rho}, the typical delay time is controlled by the quenched value of $D_\h$ itself.


In the scaling regime of the deep quench in our active system, the correlation lengths grow following power laws $\xi(t) \sim {t}^{1/z_S}$  and $\zeta(t) \sim t^{1/z_\r}$. 
\alert{By fitting the correlation lengths within the scaling regime from $tD_r=1$ to $100$, we obtain dynamic exponents $z_S=2.5 \pm 0.03$ and $z_\r=2.34 \pm 0.02$; see Fig.\ref{fig:Lc_rho_S}($c$). Over this scaling regime, correlation functions of nematic order and density show reasonable data collapse as observed in Fig.\ref{fig:Lc_rho_S}($a$) and ($b$). At shorter times, significant correlated domains have not yet formed, while at longer times, finite-size effects become noticeable.
} 


\alert{As we noted before, density and order parameters decouple in the passive limit of $Pe = 0$. As a result, a quench in $\e$ influences nematic order alone.} 
However, even estimating the growth exponent $z_S$ in 2D passive nematics is not that simple.  
For example, a soft-spin variant of the 2D Lebowhl-Lasher model estimated $z_S \approx 2.5$~\cite{Blundell1992}, but could not validate the theoretical growth law $\xi \propto (t/\ln t)^{1/2}$~\cite{bray2002theory}.
\alert{The active phase ordering exponent $z_S$ is consistent with that of the 2D equilibrium Lebwohl-Lasher model~\cite{Blundell1992}, and the growth kinetics of density domains, characterized by $z_\r$, surpass the rate predicted for model B dynamics of conserved quantities~\cite{pcmp1995}. Additionally, our findings differ from those of equilibrium model C dynamics, where coupling a conserved field with a non-conserved order parameter of $n > 1$ components renders the influence of a non-critical conserved variable unimportant, resulting in a dynamical exponent equivalent to that of model A~\cite{hohenberg1977theory, pcmp1995}. 
}


A precise determination of the growth laws and growth exponents for stochastic models of active nematic requires performing a more careful and extensive finite-size scaling procedure.

 \section{Outlook}\label{conclusion}
We presented a comprehensive phase diagram as a function of activity $\l$ and relative orientational noise $\e$ for the nematic-isotropic transition in dry active nematics with reciprocal alignment interaction. The additive nature of the interaction ensures a first-order transition via coupling of nematic order to density fluctuations, confirmed through a hydrodynamic theory and the corresponding mean-field arguments. 

It is intriguing that despite zero macroscopic velocity, the phase behavior of these apolar SPPs shows intricate dependence on $\l$. Our theoretical analysis predicts a phase boundary  $\e \sim \l^2$, aligning well with simulation results up to $\e \approx 1.4$. Beyond this, local density increases are not enough to sustain nematic order, resulting in a $\l$-independent phase boundary. The scalar order parameter shows a discontinuous jump proportional to $\l$ at the transition, disappearing in the passive particle limit of vanishing $\l$. We identified a broad parameter regime of phase coexistence. Deep within the nematic phase, the nematic bands are inherently unstable, exhibiting giant fluctuations and turbulent dynamics due to their continuous formation and breakup. 

Moreover, we investigated phase ordering kinetics by quenching from the isotropic to the nematic phase. This revealed coarsening patterns with correlation lengths in nematic order and density fluctuations growing as $t^{1/z}$ with exponents. \alert{The phase ordering exponent $z_S=2.5 \pm 0.03$ aligns with 2D equilibrium Lebwohl-Lasher model, and the exponent $z_\r =2.34 \pm 0.02$ controlling growth kinetics of density domains predicts a faster kinetics than the the rate expected for conserved quantities. Both the exponents $z_S$ and $z_\r$ remain within the bounds of model A exponent 2 and model B exponent 3,} consistent with earlier hydrodynamic predictions~\cite{mishra2014aspects}. The quench in scaled orientational coupling showed a lag in particle clustering relative to the growth of nematic correlation. 
{Our predictions are testable in experiments.}  

In conclusion, we obtained a comprehensive phase diagram for reciprocal alignment in active nematics, supported by mean-field predictions for the observed transition and phase boundaries. A theoretical description beyond the mean field explaining the full phase diagram will require further work. The numerical studies of coarsening dynamics revealed dynamic exponents with values in between the exponents for non-conserved and conserved dynamics. A precise numerical determination of growth laws requires careful system size scaling studies.  
The extent to which the observed phenomenology holds for non-reciprocal interactions warrants further investigation.

 \section*{Author Contributions}
 DC designed and supervised the research. AS performed all the numerical calculations and data analysis. DC and AS wrote the manuscript.

 \begin{figure}[t!]
\centering
 \includegraphics[width=5.5 cm]
 {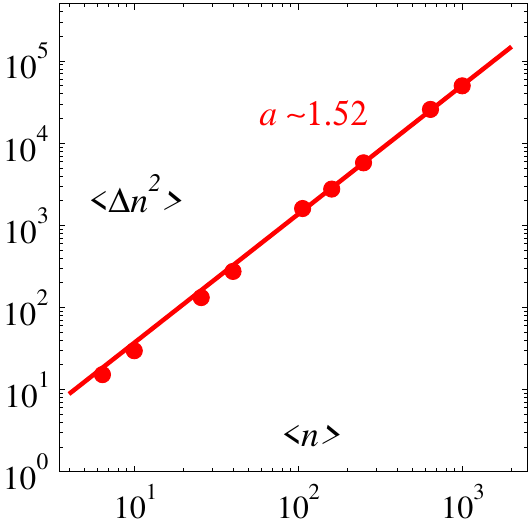}
  \caption{The number fluctuation $\la \D n^2\ra$ as a function of mean number $\la n \ra$  at $Pe=714.28$ and $\e=0.254$, calculated over square coarse graining areas of linear dimension $\ell_c/r_c=4, 5, 8, 10, 16, 20, 25, 40, 50$. A fitting, the solid line, shows $\la \D n^2\ra \sim \la n\ra^a$, with $a \approx 1.52$.   
  }
  \label{fig:gnf}
\end{figure}

\section*{Conflicts of interest}
There are no conflicts to declare.

\section*{Data availability}
The data supporting this article have been included as part of the ESI.

\section*{Acknowledgments}
The numerical simulations were performed using SAMKHYA, the High-Performance Computing Facility provided by the Institute of Physics, Bhubaneswar. D.C. thanks Sriram Ramaswamy and Abhishek Chaudhuri for useful discussions and acknowledges the Department of Atomic Energy, Government of India (1603/2/2020/IoP/R\&D-II/150288) and the International Centre for Theoretical Sciences (ICTS-TIFR), Bangalore, for an Associateship.

\appendix

\alert{
\section{Nematic order parameter in 2D}
\label{2dorder}
In two dimensions (2D), the local nematic order can be expressed as, 
\begin{equation}
 \mathbf{Q}(\rv)  = 
\f{1}{2}\begin{pmatrix}
\la \cos 2 \theta_i\ra & \la \sin 2 \theta_i\ra \\
 \la \sin 2 \theta_i\ra & -\la \cos 2 \theta_i\ra
 \end{pmatrix}
\label{Q_2d}
\end{equation}
where $\theta_i$ is the angle of the $i-$ th molecule with respect to a fixed axis. Here, the averaging $\la \dots \ra$ is assumed over a local coarse-grained volume around the position $\rv$. The traceless symmetric form of this $2 \times 2$ matrix ensures that its eigenvalues are $\pm \hf S(\rv)$ where 
 the scalar order parameter :
 \bea
S(\rv)=[\la\cos(2\theta_i)\ra^2 + \la\sin(2\theta_i)\ra^2]^{1/2}\, .
\label{S_2d}
\eea
In the eigenbasis, the order parameter takes the form
\begin{equation}
 \mathbf{Q}(\rv)  = 
\f{1}{2}\begin{pmatrix}
S(\rv) & 0 \\
0 & -S(\rv)
 \end{pmatrix} \, .
\end{equation}
Note that by definition, the nematic order parameter is bounded by $0 \leq S \leq 1$.
}

\section{Giant number fluctuations} \label{app_gnf}
 Deep inside the nematic phase, at $Pe = 714.28$ and $\e = 0.254$, we analyze the particle number fluctuations $\la \Delta  n^2\ra = \la n^2 \ra - \la n \ra^2$ as a function of $\la n\ra$; see Fig.\ref{fig:gnf}. This shows giant number fluctuations, characterized by $\la \Delta n^2 \ra \sim \la n\ra^a$, with $a \approx 1.52 > 1$ but less than $a=2$ predicted in \cite{Ramaswamy2003} and found by numerical simulations in \cite{Mishra2006, chate2006simple}. See Fig.\ref{fig:phase_diag}($g$) for a typical configuration corresponding to these parameter values.

\section{Density-nematic order cross correlation} 
\label{app_xcor}
Here, we compute the cross-correlation 
$C_{\r S}
= \la \d  \r (r)  \d S (r)  \ra/  \s_S  \s_\r$, where $\d  \r (r)= \r(r) - \la \r \ra$ and $\d S (r)=S(r) - \la S \ra$ with $\la \r \ra$ and $\la S \ra$ denoting the mean density and mean scalar order, respectively(see Fig.\ref{fig:C_rho_S}). Further, $\s_S= \la\d S^2(r)\ra^{1/2}$ and $\s_\r= \la\d \r^2(r)\ra^{1/2}$ denote the standard deviations of nematic and density fields. The calculations are performed over coarse-grained cells of linear dimension $10\,r_c$. 
In the disordered phase, $\e > 1.5$, the cross-correlation $C_{\rho S}$ is slightly negative, indicating an anti-correlation between the fields.  This happens because an increase in randomly oriented elements within a coarse-grained region can lead to better cancellation in averaging and reduce the overall nematic order. Near the transition point, $\e \approx 1.4$, $C_{\rho S}$ reaches its positive maximum because higher particle density in bands enhances order, while the low-density background remains disordered (see Fig.\ref{fig:phase_diag}($f$)\,). Deeper in the nematic phase, $\e \lesssim 0.28$, $C_{\rho S}$ decreases as regions of both high and low densities exhibit similar nematic order (see Fig. \ref{fig:phase_diag}($g$)\,).

\begin{figure}[t!]
\centering
 \includegraphics[width=5.5 cm]
 {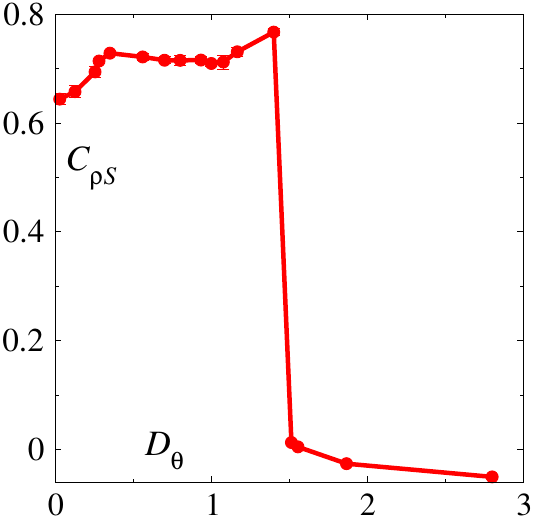}
  \caption{Cross-correlaion coefficient $C_{\r S}$ with the scaled orientation noise $\e$ at a fixed $\l=428.57$.   
 }
  \label{fig:C_rho_S}
\end{figure} 

\section{Equilibrium mean-field theory for nematic-isotropic transition}
\label{app_eqmft}
 Using Eq.\eqref{eq_LL} it is straightforward to write the mean-field Fokker-Planck equation for orientation~\cite{chepizhko2021revisiting, Sinha2024}
\bea
\p_t p (\h,t) &=& \p_\h\left[ \g (n) \int_0^{2\pi} d\h' \sin[2(\h-\h')] p(\h,t)p(\h',t) \right] \nn\\
&&+ D_r \p^2_\h p,
\label{eq_fp}
\eea
where $\g (n) = 2 \mu J n$, with $n = \pi r_c^2 \rho$ denoting the mean number of nearest neighbors. 
The steady-state solution is 
\bea
p_{st}(\h) = {\cal N} \exp\left[ \f{\g(n) S}{2 D_r} \cos(2(\h-\psi))\right]. 
\label{eq_p}
\eea
where
\bea
S= \mid \int_0^{2\pi} d\h p_{st}(\h) \exp(i\,2\h) \mid
\label{eq_s}
\eea
denotes the scalar order parameter quantifying the degree of nematic order, and $\psi$ denotes the direction of broken symmetry.

\begin{figure*}[t] 
    \centering   
    \begin{minipage}{0.25\textwidth}
        \centering
        \( (a)\, t=t_i\) \\  
\includegraphics[width=\textwidth]{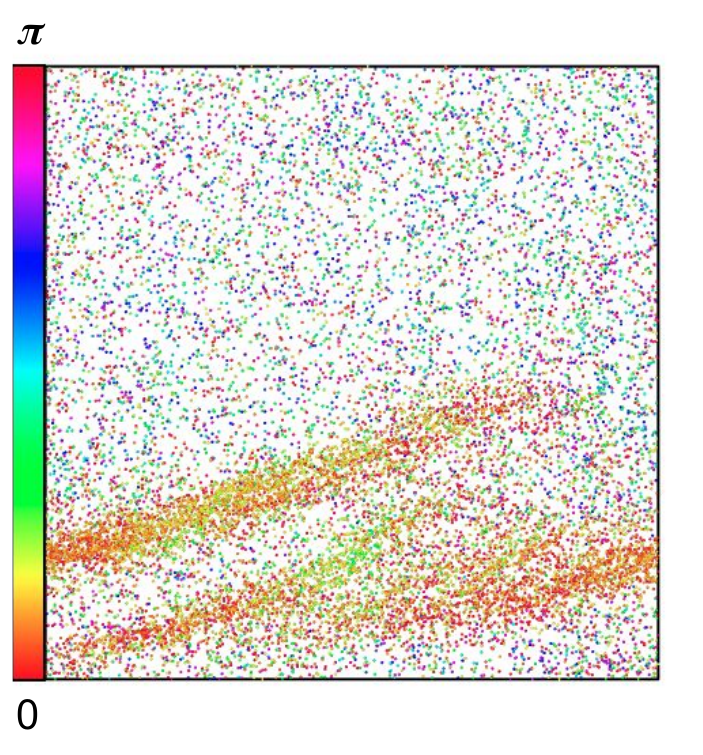}
          \end{minipage}
    \hfill
    \begin{minipage}{0.25\textwidth}
        \centering
        \( (b)\,  t_{ii}\) \\  
\includegraphics[width=\textwidth]{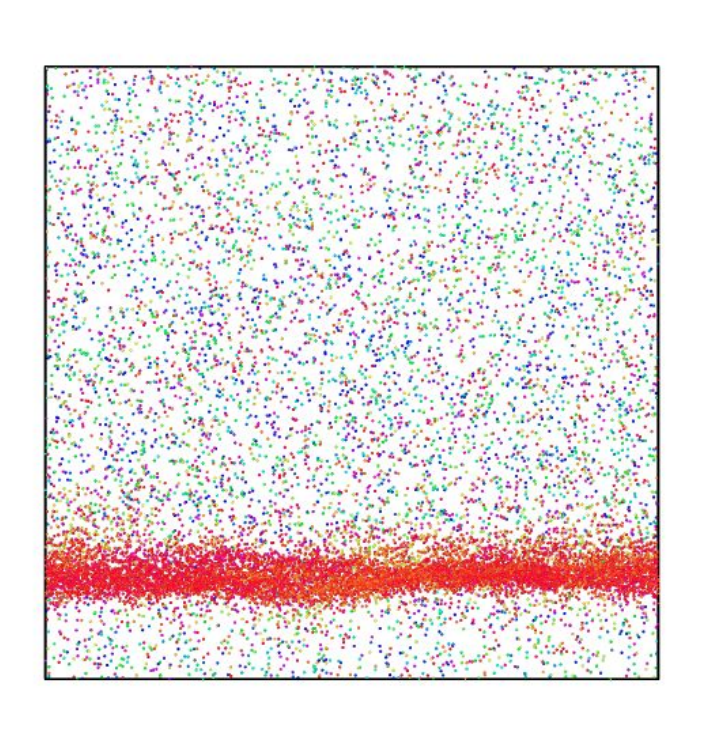}
        \end{minipage}
    \hfill
    \begin{minipage}{0.25\textwidth}
        \centering
        \( (c)\,  t_{iii}\) \\  
\includegraphics[width=\textwidth]{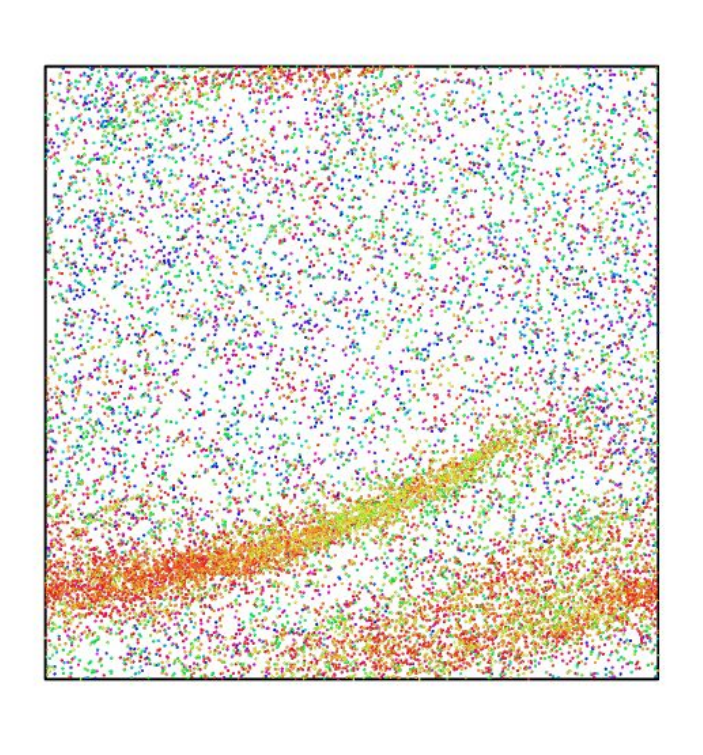}
    \end{minipage}
\caption{~Configurations of nematic bands at \((Pe, \e) = (714.28, 1.166)\): \((a)\) just before the formation of nematic band at time \(t_i = 21315\, D_r^{-1}\), \((b)\) when the band is fully formed at \(t_{ii} = 21490\, D_r^{-1}\), and \((c)\) as the band disintegrates at \(t_{iii} =21980 \, D_r^{-1}\). The color palette in (a) encodes orientations of particle heading directions. 
 }
    \label{fig:band_snap}
\end{figure*}

From equations (\ref{eq_s}) and  (\ref{eq_p}) we find the self-consistency relation $S=I_1\left[ \f{\g(n) S}{2 D_r} \right]/I_0\left[ \f{\g(n) S}{2 D_r} \right]$, where $I_n[.]$ denotes $n$-th order modified Bessel function of the first kind. For small $S$, a Taylor expansion gives, $S \approx \f{\g(n) S}{4 D_r}\left(1- \f{\g^2(n) S^2}{64 D_r^2} \right)$. Above the transition point, only one solution exists, $S=0$. Below it,
$S=(2 D_\h/D_\h^{(c)})(1-D_\h/D_\h^{(c)})^{1/2}$ where we used  $D_\h=2D_r/\mu J$
and the critical point 
\begin{equation}
    D_\h^{(c)}=n=\pi r_c^2 \rho
    \label{eq_Dh*}
  \end{equation}

The above relations can be used to obtain an approximate mean-field evolution of the scalar order $dS/dt = - \p {\cal A}/\p S$ with~\cite{pcmp1995} 
\bea
{\cal A} = -\f{\a_1}{2} S^2 + \f{\a_2}{4} S^4
\eea
where $\a_1 = 1-\f{D_\h}{\e^{(c)}}$ and $\a_2 = \f{1}{4}\left(\f{\e^{(c)} }{D_\h}\right)^2$, with $\e^{(c)} = A \r$. 

\section{Coarse-grained correlations}
\label{app_corr1}
 Using the expression of ${\bf Q}(\rv,t)$ in Eq.\eqref{eq_Qcg}, we get the following expressions for the coarse-grained quantities, 
 $
 S^2(\rv,t) = S_1^2(\rv,t) + S_2^2(\rv,t)
$
 and 
 $\tan 2\h(\rv,t) = S_2(\rv,t)/S_1(\rv,t)$
  with $S_1(\rv,t) = \la \cos 2\h(\rv ,t)\ra= \f{1}{n(t)} \sum_{i=1}^{n(t)}\cos 2\h_i$ and $S_2(\rv,t) = \la \sin 2\h(\rv ,t)\ra = \f{1}{n(t)} \sum_{i=1}^{n(t)}\sin 2\h_i$. Here, $n(t)$ denotes the instantaneous number of particles in the coarse-grained volume around $\rv$.  
 Therefore, the coarse-grained nematic correlation finally gets the form,
 \bea
C_{\bf Q}(r,t)= \f{ \la S(\rv_1,t) S(\rv_2,t) \cos 2 [\h(\rv_1, t) - \h(\rv_2,t)]\ra}{\la S^2(0,t)\ra}, \nn\\
 \eea
 where $r=|\rv_1-\rv_2|$. 
 
 The correlation lengths are obtained by fitting the numerically calculated correlation functions with exponential decay of form $\exp(-r/\ell)$ with $\ell$ denoting the correlation length. 
 Beyond the scaling regime shown in Fig.\ref{fig:Lc_rho_S}, the nematic order grows quickly to span the whole system and changes from exponential to power-law decay, characteristic of the steady-state 2D active nematic phase.  Related to this, the density correlation length also saturates. This crossover is entirely a finite-size effect, shifting to a later time and larger length scales in bigger systems. The data for this crossover is not shown explicitly in Fig.\ref{fig:Lc_rho_S}.

\section{ Nematic Bands in the coexistence region }
   \label{nematic_bands}

  In Fig.~\ref{fig:band_snap}, we present three snapshots corresponding to the parameter values \((Pe, \e) = (714.28, 1.166)\), the parameters in  Fig.~\ref{fig:phase_diag}(f) of the main text. The snapshots capture distinct moments during the simulation: \((a)\) just before nematic bands form at some time instance \(t_i = 21315 \, D_r^{-1}\), \((b)\) when the band is fully formed at 
  \(t_{ii} = t_i + 175 \, D_r^{-1}\),
  and \((c)\) as the band  disintegrates at 
\(t_{iii} = t_i + 665\, D_r^{-1}\). 
  %
  This cycle of nematic band formation and fragmentation occurs repeatedly, with the band orientations changing spontaneously between such cycles.




\bibliographystyle{rsc}


\begin{mcitethebibliography}{67}
\bibitem[]{footnote}
Even at the highest $Pe$ value of 700 used in our simulations, the condition $v_0 dt/r_c = 7 \times 10^{-4} \, Pe = 0.49$ guarantees that two particles will remain within each other's interaction radius for at least two-time steps, which equates to a total of four time steps within the interaction diameter
\EndOfBibitem
\end{mcitethebibliography}

\providecommand*{\mcitethebibliography}{\thebibliography}
\csname @ifundefined\endcsname{endmcitethebibliography}
{\let\endmcitethebibliography\endthebibliography}{}

\end{document}